\documentclass[usenatbib]{mn2e}

\usepackage{amssymb}
\usepackage{epsfig}
\usepackage{url}
\usepackage{colordvi}
\usepackage{txfonts}
\usepackage{graphicx}
\usepackage{bm}

\def\araa{ARA\&A}%
\def\aap{A\&A}%
\newcommand{\bez}{\begin{eqnarray*}}
\newcommand{\eez}{\end{eqnarray*}}
\newcommand{\be}{\begin{equation}}
\newcommand{\ee}{\end{equation}}
\newcommand{\beq}{\begin{eqnarray}}
\newcommand{\eeq}{\end{eqnarray}}
\newcommand{\bc}{\begin{center}}
\newcommand{\ec}{\end{center}}

\def\msun{{\rm M_{\odot}}}

\def \pd{{\partial}}
\def\nut{\nu_{\rm t}}
\newcommand{\frad}{\mbox{\boldmath ${\cal F}_{\rm rad}$}}
\newcommand{\fradscal}{{\cal F}_{\rm rad}}
\newcommand{\SSSS}{\mbox{\boldmath ${\sf S}$} {}}

\title[Boundary layer on the surface of a neutron star]
{Boundary layer on the surface of a neutron star}

\author[Babkovskaia et al.]
{N.\ Babkovskaia,$^{1,2}$\thanks{E-mail: nbabkovs@nordita.org}
A.\ Brandenburg$^2$ and J.\ Poutanen$^1$\\
$^1$  Astronomy Division,  Department of Physical Sciences, P.O. Box 3000,
FIN-90014 University of Oulu, Finland\\
$^2$NORDITA, Roslagstullsbacken 23, AlbaNova University Center,
106 91 Stockholm, Sweden
}

\begin{document}

\date{Accepted 2008 February 12. Received 2008 January 29; in original form 2007 November 16}

%\date{\today}

\pagerange{\pageref{firstpage}--\pageref{lastpage}} \pubyear{2008}

\maketitle \label{firstpage}

\begin{abstract}
In an attempt to model
the accretion onto a neutron star in low-mass X-ray
binaries, we present two-dimensional hydrodynamical models of the
gas flow in close vicinity of the stellar surface.
First we consider a gas pressure dominated case, assuming that the
star is non-rotating. For the stellar mass we take $M_{\rm star}=1.4
\times 10^{-2} \msun$ and for the gas temperature $T=5 \times
10^6$ K. Our results are qualitatively different
in the case of a realistic neutron star mass and a realistic gas
temperature of $T\simeq 10^8$ K, when the radiation pressure
dominates.  We show that to get the stationary solution in a
latter case,  the star most probably  has to rotate with the
considerable velocity.
\end{abstract}


\section{Introduction}

Low-mass X-ray binaries (LMXB) are luminous X-ray sources composed
of a late-type optical companion (mass less than about 1 solar
mass) and a neutron star. About 100 low-mass X-ray binaries are
known now. Neutron stars in such objects are most probably old and
have a rather weak magnetic field so that an accretion disk can extend down
to the neutron star surface. The rapidly rotating gas is
decelerating due to viscous friction. The gas then  spreads over
the stellar surface and forms a  boundary layer. Here most of the
energy is emitted in the form of X-rays, whilst its amount
is comparable with the energy generated in the accretion disk
\citep{ss86,ss98}.

LMXBs can be divided into two different classes. Very luminous
Z-sources ($L\sim$ 0.1 $-$ 1$L_{\rm edd}$) have relatively soft,
two-component spectra, while both components can be approximated
by black bodies  with color temperatures of about
1~keV and 2.5~keV, respectively \citep{GRM03}. The other less luminous sources ($L\sim$
0.01 $-$ 0.05$L_{\rm edd}$) are observed in two states: the
high/soft and low/hard states. The radiation spectra in the soft
state are similar to those of the Z-sources, while in the hard
state they are close to the spectra of the Galactic black holes in
the hard states \citep{barret00}.

The soft component can be associated with  the radiation
from the accretion disk, while the hard one is produced in
the boundary layer \citep{mitsuda84,GRM03}. On the other hand, the
spectra from the spreading layer depend on the neutron star
compactness (mass/radius ratio), which determines the gravitational
field at the surface. Therefore, one can get independent
constraints on the equation of state of the matter at extreme
densities, calculating the spectra and comparing them with the
observational data \citep{sp06}.

A study of the motion of the matter very close to the neutron star
is also important for understanding the production of
quasi-periodic oscillations  (QPOs) observed in the kHz range from
a number of accreting neutron stars in LMXBs \citep{vdk00}.
These QPOs may
provide direct ways of measuring effects that are unique to the
strong gravitational-field regime. However, the question about the
nature of QPOs is still open, partly because of the complexity of
hydrodynamical flows in close vicinity of a neutron star
surface. Thus, detailed studies of the structure of the boundary
layer plays the key role in understanding  the physics in the vicinity
of a compact object.

The first model of the boundary layer was proposed by
\citet{pringle77}, who considers it as part of the
accretion disk. In his model the gas is moving  parallel to the
disk mid-plane and is decelerating due to differential rotation
and viscous forces. The effective temperature of the boundary
layer appears to be higher than the maximum accretion disk
temperature, because the size of the BL is smaller than that of
the disk, whereas their luminosities are comparable. \cite{PN92}
identified non-physical aspects of the standard $\alpha$-viscosity
prescription and developed a more physically realistic model of
viscosity. \citet{NP93} proposed a self-consistent  model of  a
boundary layer on the surface of a white dwarf and accounted for
the hard X-rays observed in cataclysmic variables.
\citet{medvedev04} studied the radiative accretion onto a rapidly
spinning neutron star. They considered a quasi-spherical hot
settling accretion flow and presented an analytical self-similar
solution describing the boundary layer.

\citet{is99}  considered the boundary layer as a
spreading layer on the surface of the neutron star. They proposed
that matter spirals along the neutron star surface toward the
poles due to turbulent friction between matter and stellar
surface.  They used a 1D approach, averaging all values in the
radial direction and assuming azimuthal symmetry. To describe the
turbulent viscosity they used the Prandtl-Karman universal
logarithmic dependence of the mean velocity on the distance from
the stellar surface and introduced a turbulent viscosity to
characterize turbulent velocity and turbulent pressure fluctuations.
They also assumed
that the dissipation of rotational kinetic energy causes a strong
energy release near the bottom of the boundary layer.
With these simplifications they constructed a semi-analytical model
and showed that the kinetic energy of the gas is mostly liberated in two
belts above and below the equator of the neutron star.

In order to solve the 1D spreading layer problem,
\citet{is99}   assumed that the initial rotational velocity in the
equatorial plane is very close to Keplerian.
However, because of the presence of the boundary layer associated with the accretion disk,
this velocity can significantly deviate from the Keplerian value.
There is no doubt that
the behavior of the gas at higher  latitudes in the spreading layer strongly depends on
the conditions in the equatorial plane.
Therefore it is important to describe the gas
flow near the equatorial point more accurately.
As a first step we present here two-dimensional numerical hydrodynamic
solutions in the neighborhood of the equatorial point.

\section{Equations and coordinates}

The full  non-stationary system of the hydrodynamical equations is
as follows. The continuity equation is solved in the form
\begin{eqnarray}
\frac{{\rm D}\ln \rho}{{\rm D} t} = -{\bm\nabla} \cdot \bm{U},
\label{eq:rho}
\end{eqnarray}
where $\rho$ and ${\bm U}$ are density and velocity of the gas,
and ${\rm D}/{\rm D}t= \pd/\pd t + {\bm U}\cdot {\bm \nabla}$ is the advective
derivative. The conservation of momentum can be written in the
form
\begin{eqnarray}
\frac{{\rm D} {\bm U}}{{\rm D} t}= -\frac{1}{\rho} \nabla p +{\bm
F_{gr}}+{\bm F_{vs}} +\frac{\kappa \frad}{c},
\label{eq:UU}
\end{eqnarray}
where $\bm{F}_{\rm gr}=-GM_{\rm star}{\bf r}/r^3$ is the
gravitational force (where $M_{\rm star}$ is a stellar mass, $\bf
r$ is a radius-vector), $\bm{F}_{\rm vs}=\rho^{-1}{\bm
\nabla}\cdot(2\nut \SSSS)$ is the viscous force, $\nut$ is the
turbulent viscosity, $\frad$ is a radiative flux, $\kappa$ is the
opacity, and $c$ is the speed of light. The energy equation is
formulated in terms of specific entropy $s$,
\begin{eqnarray}
T \frac{{\rm D} s}{{\rm D} t}  = 2 \nut \SSSS^2 - \frac{1}{\rho} {\bm
\nabla} \cdot \frad , \label{eq:entropy}
\end{eqnarray}
where $T$ is the temperature and
$\SSSS  ={1\over2}(U_{i,j}+U_{j,i})-{1\over3}\delta_{ij}\nabla\cdot\bm{U}$
is the trace-less rate of strain  tensor
and commas denote partial differentiation.
In the following we assume that the gas is optically thick and can
therefore treat radiation in the diffusion approximation, so the
radiative flux is given by
\begin{eqnarray}
\frad=-K \bm{\nabla} T,
\end{eqnarray}
where $K=16\sigma_{\rm SB}T^3/(3\kappa\rho)$ is the radiative conductivity.

\begin{figure}
\centerline{\epsfig{file=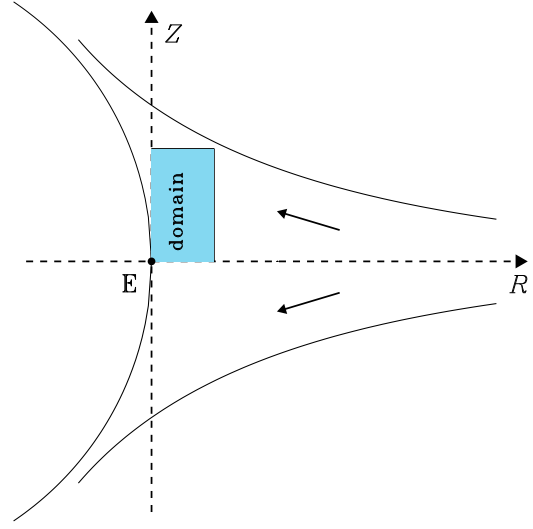,width=7cm}} \caption{Sketch of
the boundary layer on the surface of the neutron star.}
\label{fig:geom}
\end{figure}

\begin{figure}
\centerline{ \epsfig{file=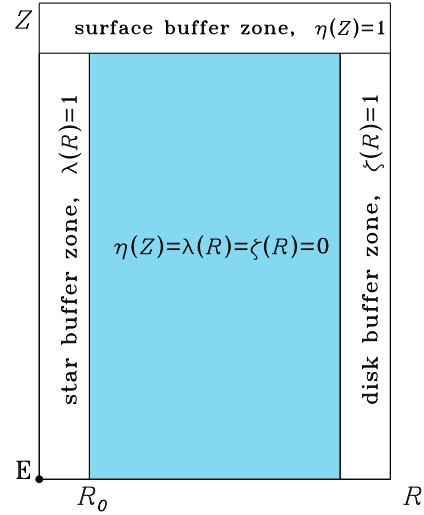,width=7cm}} \caption{Sketch
of the calculated domain and of the buffer zones.}
\label{fig:domain}
\end{figure}

A sketch of the boundary layer on the surface of the neutron star
is presented in Fig.~\ref{fig:geom}. The gas is accreting in the
disk mostly in the radial direction $R$ and turns in the
$Z$-direction near the equatorial point $E$. Since the purpose  of
this paper is to study the gas flow in the vicinity of $E$, we use
cylindrical coordinates, and neglect the curvature of the stellar surface.
We consider a 2D domain limited in the radial direction by the surface of the
star and the disk zone, where the rotational velocity $U_{\Phi}$
is close to the Keplerian value $U_{K}$ (see Fig.~\ref{fig:domain}).

\begin{table}
\begin{center}
\caption{Parameters}
\begin{tabular}{cccc}
quantity &  case~1 & case~2 \\
\hline
 $R_{\rm star}$  &$10^{6}$ cm & $10^{6}$ cm & \\
 $ M_{\rm star}$ & $ 1.4 \times 10^{-2}$ M$_\odot$ &  1.4  M$_\odot$ &\\
 $T_{\rm star}$&  $3\times10^{6}$ K &  $10^{8}$ K &\\
 $T_{\rm disk}$&  $1.5\times 10^{6}$ K & $10^{8}$ K & \\
 $\rho_{\rm disk}(0)$& $0.1$ g cm$^{-3}$ &$4$ g cm$^{-3}$ & \\
 $\nut$ & 10$^{7}$, 10$^{8}$ cm$^2$ s &10$^{10}$ cm$^2$ s$^{-1}$ & \\
 $\alpha$  & 0 & 1 & \\
\hline

\end{tabular}
\label{tab:parameters}
\end{center}
{$R_{\rm star}$ is a stellar  radius, $M_{\rm star}$ is  stellar
mass, $T_{\rm star}$ and $T_{\rm disk}$ are temperatures of the
star surface and the gas in the disk zone, $\rho_{\rm disk}(0)$ is
the gas density at the mid-plane in the disk zone,  $\nut$ is a
turbulent viscosity.}
\end{table}

\section{Boundary  conditions and buffer zones}

The boundary conditions represent an integral part of the
overall solution in that the values on the boundaries both determine and
depend themselves on the final solution.
They must allow the accretion onto the stellar
surface and the emission of energy.
They must also simulate the compression
of gas near the surface and allow this gas to
become part of the surface, while the gas settling depends
itself on the input parameters of incoming gas and conditions on
the stellar surface.
The rotational velocity of the gas in the disk
part is close to Keplerian, but cannot be exactly Keplerian, because
otherwise accretion would stop and the boundary layer would
disappear. On the other hand, the deviation from the Keplerian
velocity is determined by the conditions on the stellar surface.
To cope with these complications we use so-called buffer zones, which are
narrow regions just outside the domain (of the size of typically 5 grid-points;
see Fig.~\ref{fig:domain}), where
additional terms are added to the hydrodynamical equations.
This type of approach has proven to be useful in earlier simulations
of disk outflows and star-disk coupling \citep{vr03}.
We use three different buffer zones that are characterized by
the three profile functions $\zeta(R)$ for the disk buffer zone,
$\lambda(R)$ for the star buffer zone, and
$\eta(Z)$ for the surface buffer zone.
In the following we describe the properties of these three zones separately.

In the disk buffer zone, where $\zeta(R)=1$, the gas is
accelerated close to the Keplerian speed due to an additional
source term in the equation for $U_\phi$, while $U_R$ adjusts itself to
the conditions inside the domain.
Thus, the $\phi$ and $R$ components of equation
(\ref{eq:UU}) are modified by additional terms
on their right hand sides,
\begin{eqnarray}
&&\frac{{\rm D} U_{\Phi} }{{\rm D} t}=...-\frac{U_{\Phi}-U_K(R)}{\tau} \zeta(R),
\\
&&\frac{{\rm D} U_{R}^{j}}{{\rm D}
t}=...-\frac{U_{R}^j-U_{R}^{j-1}}{\tau}\zeta(R),\quad j=1, ...,
N_R, \label{eq:velosity_1}
\end{eqnarray}
where $j$ denotes the meshpoint in the $R$ direction,
$N_R$ is total number of the grid points in the $R$-direction,
and dots indicate the presence of terms that where already specified
in equation (\ref{eq:UU}).
We take $\tau=5\delta t$, where $\delta t$ is the length of the time step.

In the buffer zone near the star, where $\lambda(R)=1$, the radial
gas velocity goes down to zero at the stellar surface. To describe
the rotation of the star we introduce the parameter
$0~\le~\alpha~\le~1$, which equals zero if the star is
non-rotating one, and unity if it rotates with the corresponding
Keplerian velocity. Thus the $\phi$ and $R$ components of equation
(\ref{eq:UU}) are modified further by the terms
\begin{eqnarray}
&&\frac{{\rm D} U_{\Phi} }{{\rm D} t}=...-\frac{U_{\Phi}-\alpha U_K}{\tau}\lambda(R),
\\
&&\frac{{\rm D} U_{R}}{{\rm D} t}=...-\frac{U_{R}}{\tau}\lambda(R),
\label{eq:velosity_2}
\end{eqnarray}
where $\lambda(R)=1$ in the buffer zone, and zero outside.

The surface buffer zones will be discussed separately in the following
two sections, because they have to be treated differently for gas
and radiation pressure dominated regimes.

 The temperatures on the stellar surface
and the disk (left and right boundaries of the domain) are
fixed by $T_{\rm star}$ and $T_{\rm disk}$, respectively, while the gas density
is extrapolated on both sides. On the lower boundary of the domain
(mid-plane of the disk) we use antisymmetric boundary
conditions for the $Z$-component of the velocity and symmetric
boundary conditions for all other quantities, while on the upper
domain boundary all quantities are extrapolated.   The turbulent
viscosity $\nut$ is assumed to be constant everywhere in the
domain.

For all simulations presented here we use the \textsc{Pencil
Code},\footnote{\url{http://www.nordita.dk/software/pencil-code}}
which is a high-order finite-difference code (sixth order in space and
third order in time) for solving the compressible hydrodynamic equations
\citep{bd02}.

\section{Gas pressure dominated case}
\label{sec:gpd}

As a first test we consider a gas pressure dominated case and
choose the gas temperature in the disk to be $T_{\rm disk}=1.5 \times
10^6$ K. This means that the radiation pressure is about two
orders of magnitude smaller than the gas pressure. Also, we take the
stellar mass to be $M_{\rm star}= 10^{-2}\msun$ so as to balance the
gravity force near the surface by the gas pressure force.
In addition, we assume that the star does not rotate ($\alpha=0$).
We consider two cases with $\nut =10^7$ and $10^8$~cm$^2$ s$^{-1}$.

Since initially the disk is assumed to be in vertical hydrostatic
equilibrium, the vertical velocity should be close to zero.
In addition, we let the gas density $\rho$ approach a certain
vertical profile $\rho_{\rm disk}(Z)$, where the value at the disk
mid-plane is $\rho_{\rm disk}(0)=0.1$ g cm$^{-3}$, and assume that
$\rho$ decreases exponentially with $Z$. However, since the
temperature profile results from a thermal balance between
viscous heating and radiative cooling, the local sound speed
$c_{\rm s}$ in Eq.~(\ref{eq:dens}) is recalculated at each time
step. This allows the vertical density profile to adjust to the
conditions inside the domain. Thus, we have in the disk buffer zone
\begin{eqnarray}
&&\frac{{\rm D} \ln \rho}{{\rm D} t}=...-\frac{\ln\rho-\ln\rho_{\rm
disk}}{\tau}\,\zeta(R),
\label{eq:dens}\\
&&\frac{{\rm D} U_Z}{{\rm D} t}=...-\frac{U_Z}{\tau}\,\zeta(R),
\end{eqnarray}
where
\begin{equation}
\rho_{\rm disk}(Z)=\rho_{\rm
disk}(0)\exp\left(-{Z^2\over2H^2}\right),\quad
\mbox{and}\quad{1\over H^2}={\gamma GM_{\rm star}\over R^3 c_{\rm
s}^2},
\end{equation}
where $\zeta(R)=1$ in the disk buffer zone, and zero outside.

In the surface buffer zone, where $\eta(Z)=1$, we assume vanishing
first derivatives for all three velocity components $U_i$ ($i=1,
..., 3$) and for the specific entropy.
We also correct the density profile to account for the
resulting artificial pressure force which works
against the vertical gravity. Due to this term, the gas flows out
through the surface boundary rather than coming into the domain at the
beginning of the calculation.
Thus, we add the terms
\begin{eqnarray}
&&\frac{{\rm D} \ln \rho^j}{{\rm D} t}=...-\frac{\ln\rho^j-\ln\rho^{j-1}
+Z^2/(2H^2)}{\tau} \eta(Z_{j-1}),
\\
&&\frac{{\rm D} U^j_i}{{\rm D} t}=...-\frac{U_i^{j}-U_i^{j-1}}{\tau}
\eta(Z_j),
\\
&&\frac{{\rm D} s^j}{{\rm D} t}=...-\frac{s^{j}-s^{j-1}}{\tau} \eta(Z_j), \;
j=1, ..., N_R.
\end{eqnarray}

\begin{figure*}
\begin{center}
\leavevmode \epsfxsize=7.5cm \epsfbox{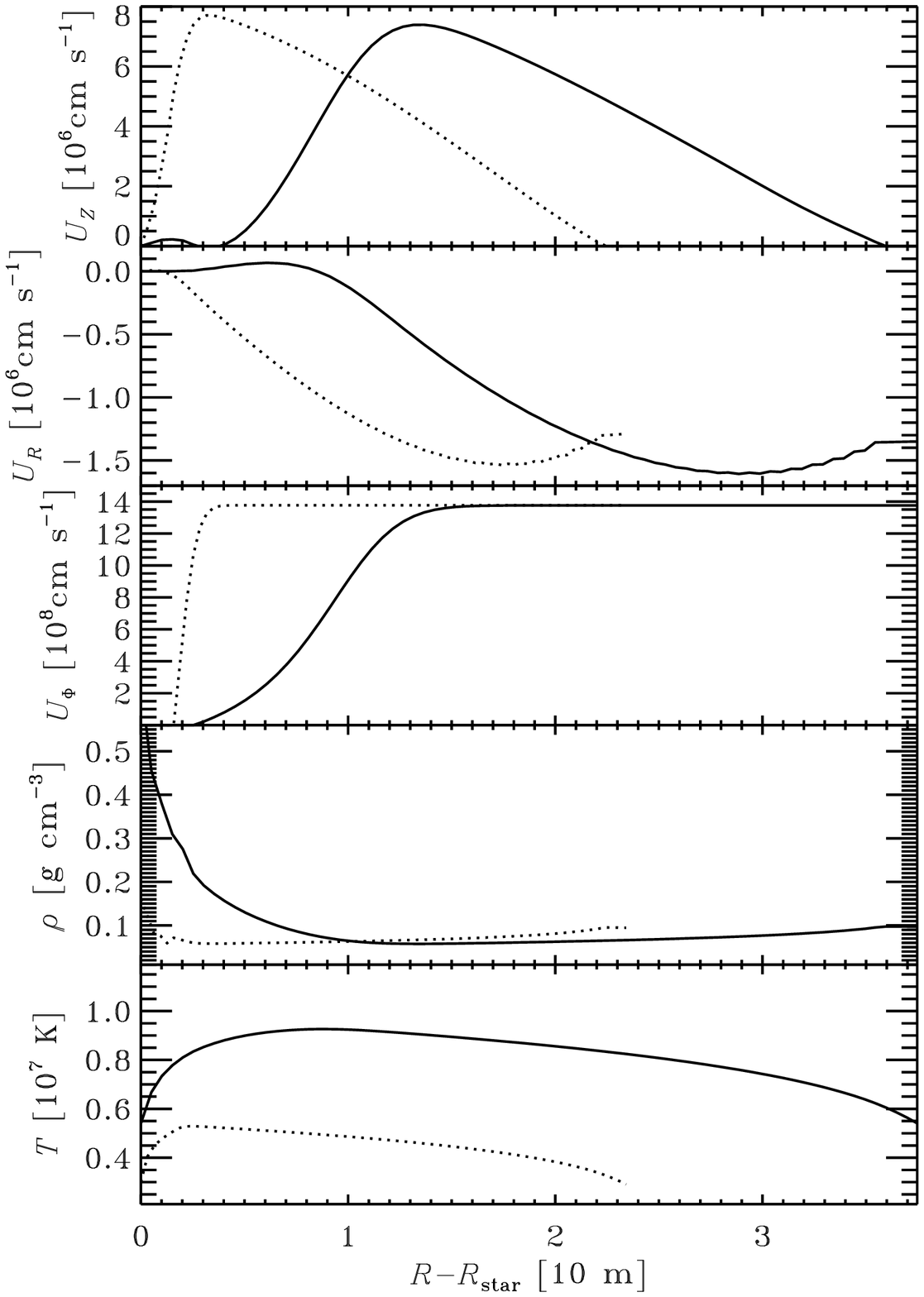} \hspace{1cm}
\epsfxsize=7.5cm \epsfbox{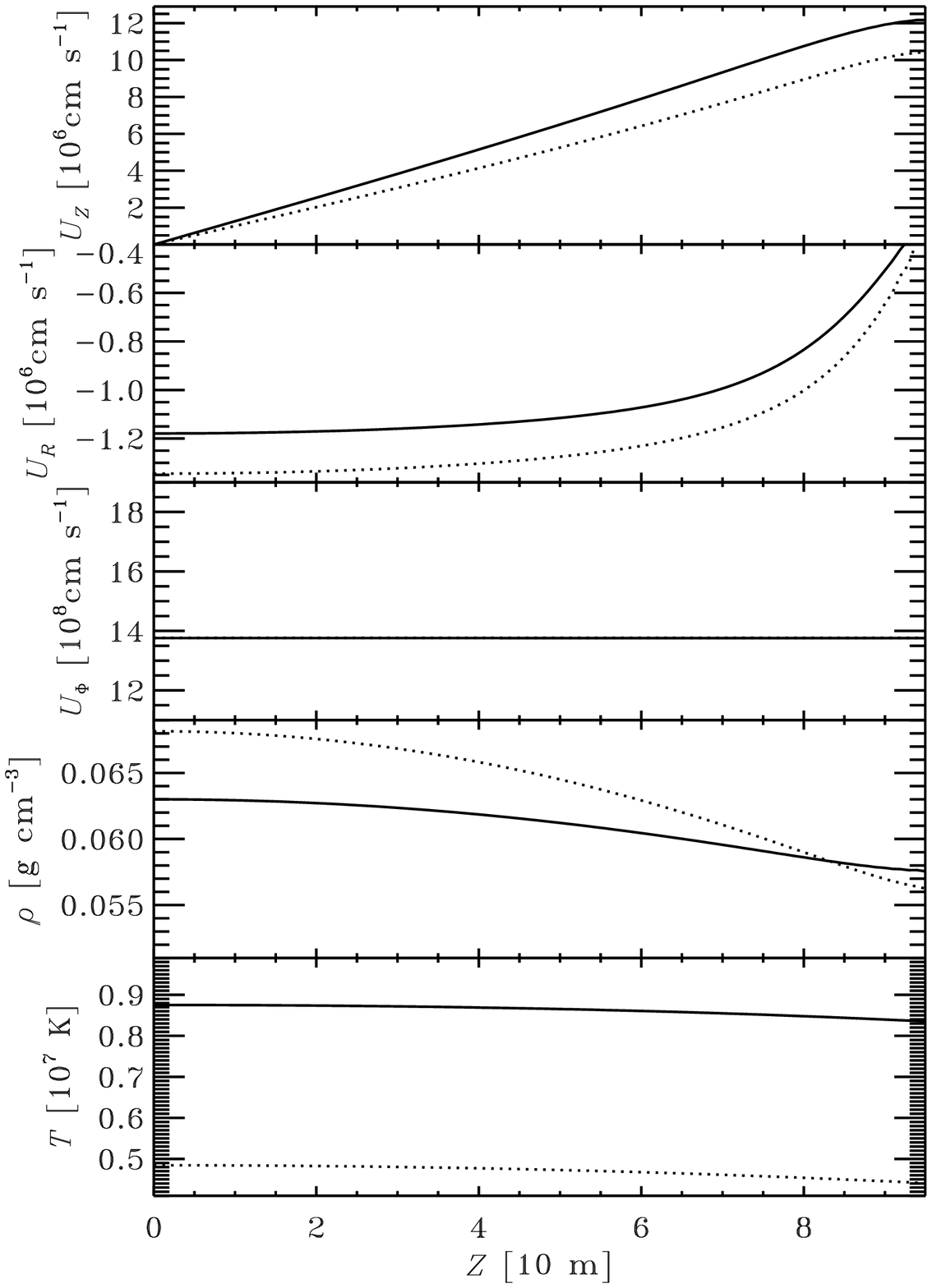}
\end{center}
\caption{Gas velocity, density and temperature as a function of
$R$ and $Z$ for the gas pressure dominated case ($M_{\rm star}=
1.4 \;10^{-2}\msun$). The dotted and solid lines correspond to
$\nut=10^7$ and $10^8$ cm$^2$ s$^{-1}$, respectively. {\it Left
panel}:  fixed $Z=50$ m, {\it right panel}: fixed $R-R_{\rm
star}=12.5$ m   for $\nut=10^7$ and $R-R_{\rm star}=20$ m  for
$\nut=10^8$ cm$^2$ s$^{-1}$ cases.} \label{fig:result1_1}
\end{figure*}

\begin{figure}
\centerline{\epsfig{file=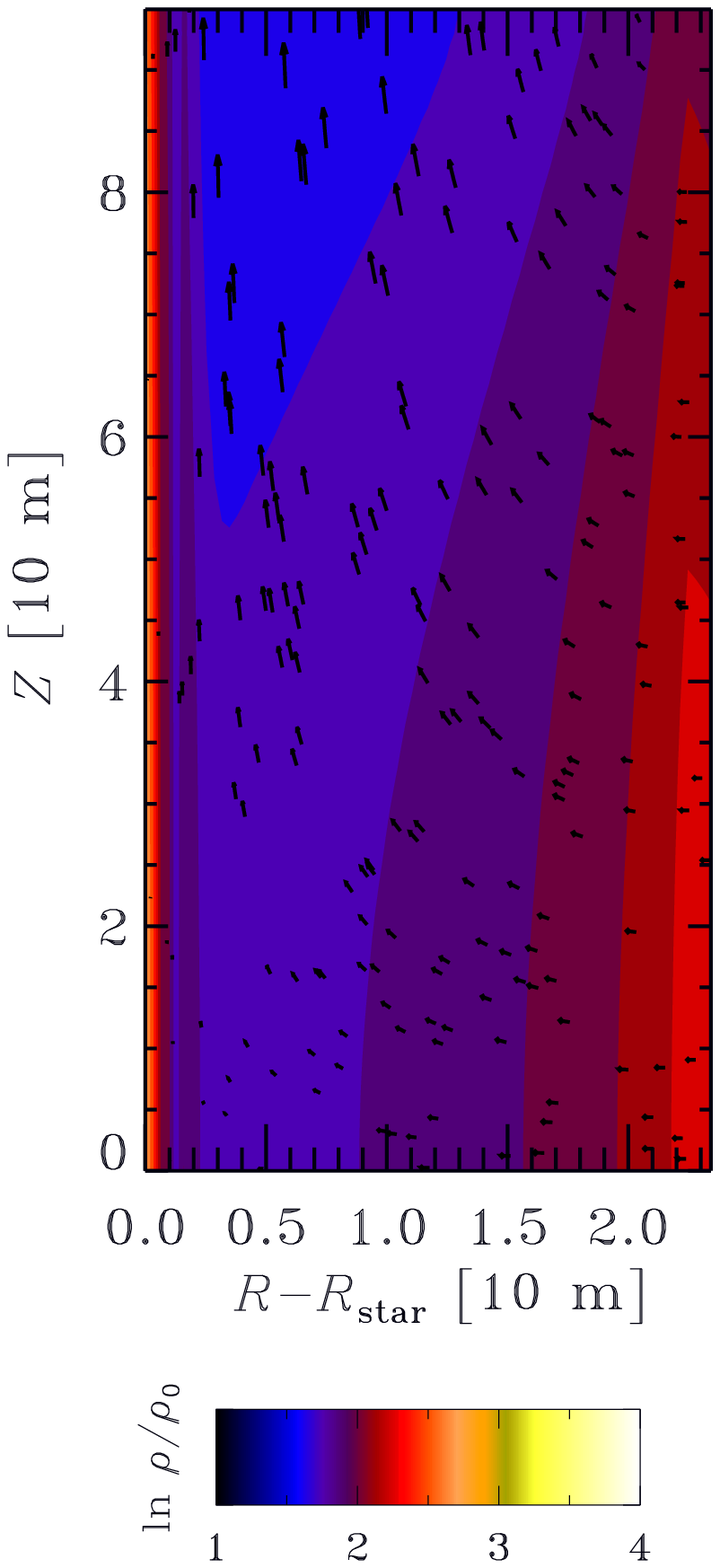,width=7cm}} \caption{Density and
velocity fields in a vicinity of a star in a gas pressure
dominated case ($M_{\rm star}= 1.4 \;10^{-2}\msun$). The domain is
limited in the radial direction by the surface of a neutron star
and in the disk midplane. The surface and disk buffer zones are
excluded. The viscosity is $\nut =10^{7}$ cm$^2$ s$^{-1}$, the
density scale is $\rho_0=10^{-2}$ g cm$^{-3}$}.
\label{fig:result1}
\end{figure}

We consider two runs with $\nut=10^7$ and $10^8$ cm$^2$ s$^{-1}$
and show in Fig.~\ref{fig:result1_1} two cross-sections respectively for
$R-R_{\rm star}=1.25$~m and 2~m, $\nut=10^7$ cm$^2$ s$^{-1}$ and
$10^8$ cm$^2$ s$^{-1}$, and $Z=5$~m in both cases.
In Fig.~\ref{fig:result1} we show velocity and density.
One can see that the accreting gas
comes to the stellar surface and turns toward the poles of the neutron
star.

We find that the size of the boundary layer, where the
rotational velocity $U_{\phi}$ of the gas decreases from the Keplerian value
down to zero, strongly depends on the value of the turbulent
viscosity. The boundary layer becomes  3.5 times thicker if one
increases $\nut$ by a factor of 10. This is consistent with the
classical theory of a boundary layer, according to which the
thickness of the boundary layer is inversely proportional to
the square root of the Reynolds number, $\sqrt{Re} \sim 1/\sqrt \nut$
\citep[see, for example, ][]{sip62}. The increase of viscosity
also leads to a growth of the gas temperature, resulting from a
balance between turbulent viscous friction and radiative
cooling. One can see that the temperature achieves its maximal
value in the middle of the boundary layer, where the velocity
gradient and therefore the heating rate are maximum.

Note, that the solution for our test case looks similar to the spreading layer
model for the white dwarf case.
We find that the $Z$-component of the gas velocity $V_Z=10^6$ cm s$^{-1}$
is very close to that obtained by \citet{pb04}.
Unfortunately, we cannot compare other quantities because
the values in the spreading layer model are
averaged along the $R$-direction.

\section{Radiation pressure dominated case}

\begin{figure}
\centerline{\epsfig{file=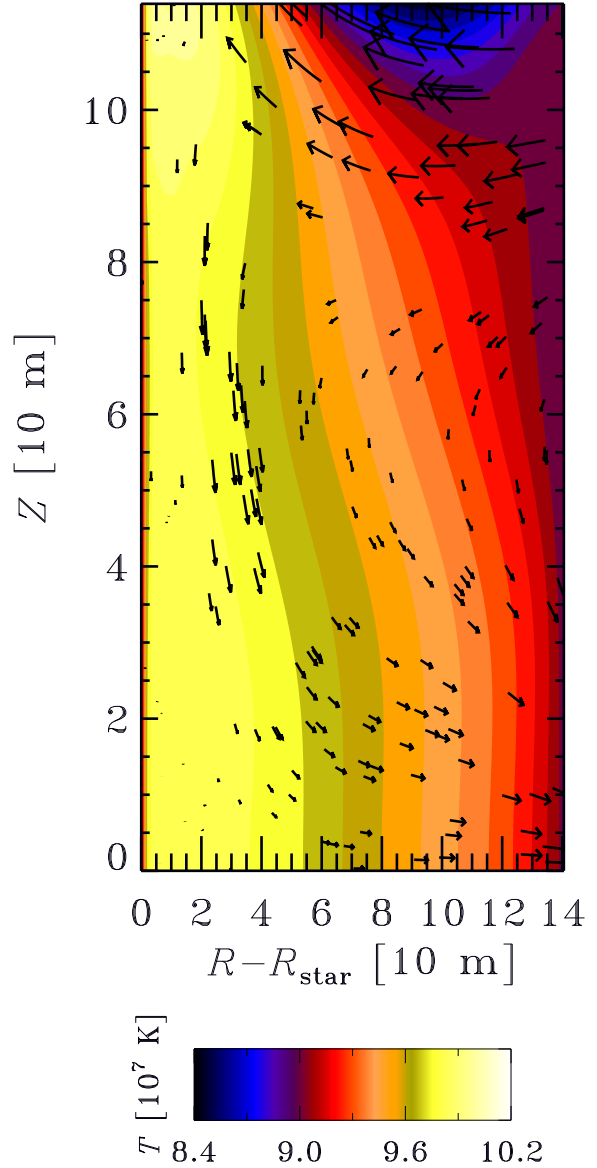,width=7cm}}
\caption{Temperature and velocity fields in a vicinity of a
neutron star in a radiation pressure dominated case ($M_{\rm
star}= 1.4\;\msun$). The domain is the same as in
Fig.~\ref{fig:result1} for $\nut =10^{10}$ cm$^2$ s$^{-1}$.}
\label{fig:result2}
\end{figure}

\begin{figure}
\centerline{\epsfig{file=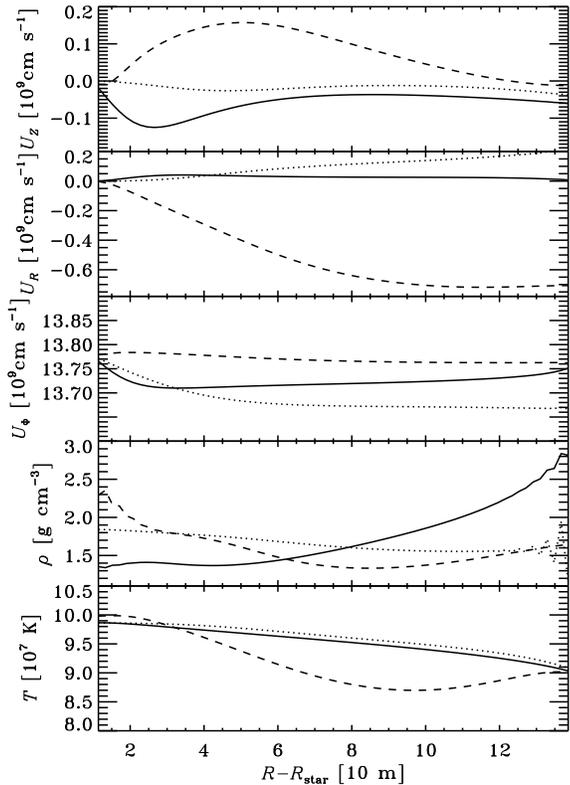,width=8cm}} \caption{Gas
velocity, density and temperature as a function of $R$ for the
fixed $Z=10$ m (dotted curve), $Z=50$ m (solid curve) and $Z=110$
m (dashed curve). The star buffer zone is excluded.}
\label{fig:result2_2}
\end{figure}

\begin{figure}
\centerline{\epsfig{file=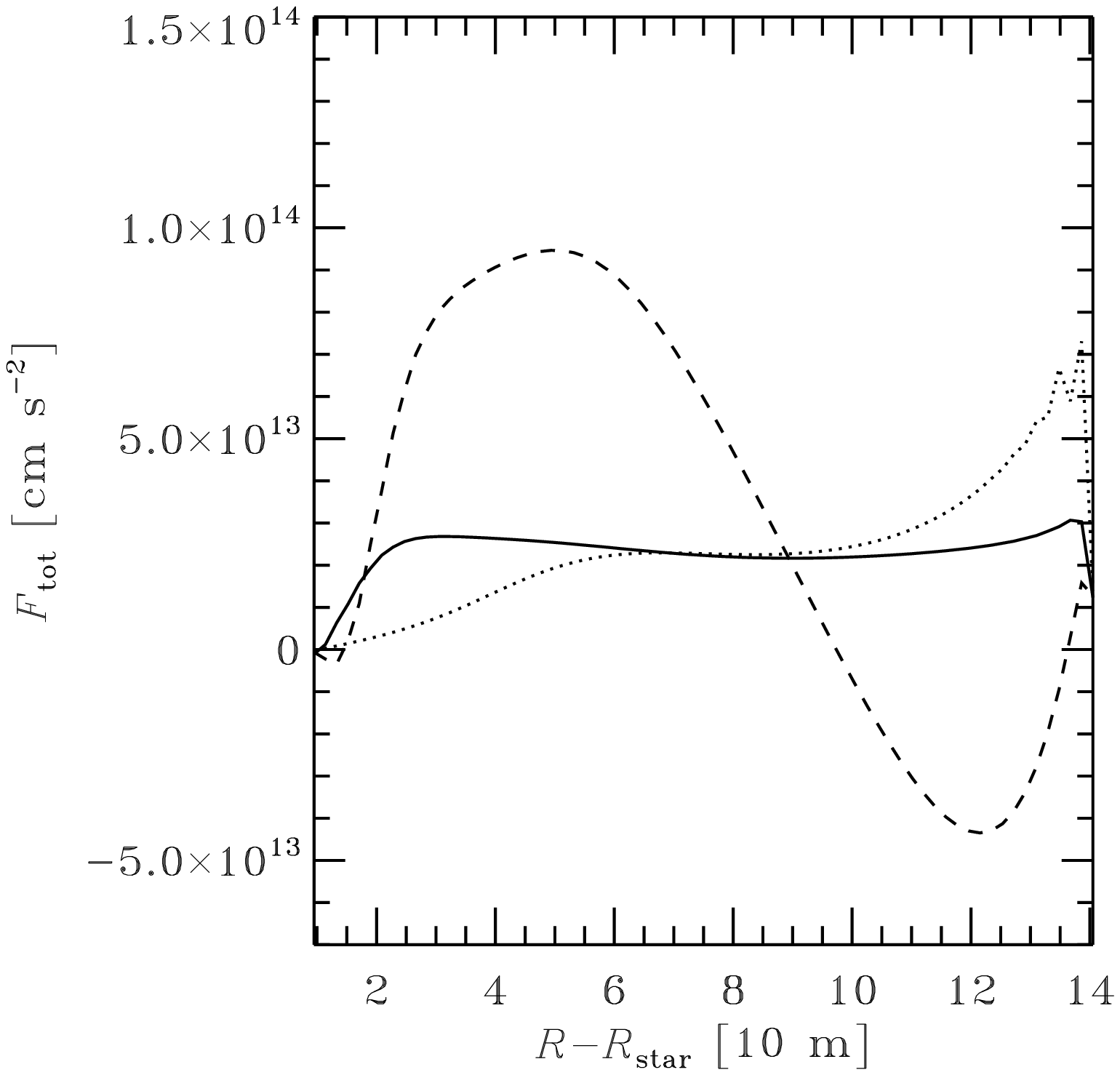,width=8cm}} \caption{The sum
of the gravitational, centrifugal and radiation pressure forces in
the $R$-direction as a function of $R$ for values of $Z$ as in
Fig.~\ref{fig:result2_2}} \label{fig:forces}
\end{figure}

Let us now consider a realistic neutron star mass, $M_{\rm
star}=1.4\msun$. We find that, in order to balance the gravity
force by the gas pressure gradient near the stellar surface, the
gas temperature must attain a value of $3\times 10^{12}$ K, which
is unrealistic. Therefore, in the case of a non-rotating star, the
only force which can work against gravity is the radiation
pressure gradient. Indeed, if one takes the gas temperature to be
$T= 3 \times 10^8$ K and $\rho \gtrsim 0.3$ g cm$^{-3}$, the
radiation pressure force becomes comparable to the gravitational
force (i.e.\ $GM_{\rm star}/R^2 \lesssim  \sigma_{\rm SB} T_{\rm
star}^4/(c \rho)$, where $\sigma_{\rm SB}$ is the
Stefan-Boltzmann constant).

Using standard  disk theory, we estimate the  turbulent
viscosity near the stellar surface $\nut =\alpha_{\rm d} H c_{\rm s}\simeq
10^{10}$ cm$^{2}$ s$^{-1}$, where $\alpha_{\rm disk} \simeq 0.01$ is a
viscosity  parameter, $H\simeq 0.01 R_{\rm star}$, and $c_{\rm s}\simeq
10^8$ cm s$^{-1}$. Note, that the radiative viscosity $\nu_{\rm r}=4
\sigma_{\rm SB} T^4 m_{\rm p}/(\kappa c^2 \rho) \simeq 10^8$ cm$^{2}$
s$^{-1}$ (where $m_{\rm p}$ is a proton mass) is much smaller than the
turbulent one, and can be neglected.

We find that  the description of buffer zones must be modified
in the radiation pressure dominated case.
For simplicity we use a similar density profile in the
disk buffer zone as it was in the gas pressure dominated case.
However, we now take $\rho_{\rm disk}(0)=4 $ g cm$^{-3}$ and fix
the gas temperature $T_{\rm disk}$ to avoid large radiation
pressure gradients and therefore the generation of large velocities,
which lead to strongly non-stationary behavior in the buffer zone and
eventually to numerical instability. Also, we use a softer
condition for the $Z$-velocity by assuming vanishing first
derivatives. Thus, we set
\begin{eqnarray}
&&\frac{{\rm D} \ln \rho}{{\rm D} t}=...-\frac{\ln\rho-\ln\rho_{\rm
disk}}{\tau}\,\zeta(R),
\label{eq:dens_2}\\
&&\frac{{\rm D} U_Z^j}{{\rm D} t}=...-\frac{U_Z^j-U_Z^{j-1}}{\tau}\,\zeta(R),
\; j=1, ..., N_R
\\
&&\frac{{\rm D} s}{{\rm D} t}=...-\frac{T/T_{\rm disk}-1}{\tau}\,\zeta(R).
\end{eqnarray}

In the course of the calculation we find that inside the domain
cold low-density patches surrounded by denser hot gas appear
sporadically. Such patches are dispersed due to motion of the gas
from the core of the patch outward through the radiation pressure
gradient. However, if we were to fix the $Z$ and $R$ velocities in
the star buffer zone to zero, such a patch cannot disappear, while
the density inside this patch is going to decrease together with
the temperature, and a numerical instability develops. To avoid
this we use symmetry conditions relative to the inner boundary of
the buffer zone $R=R_0$ (see Fig.~\ref{fig:domain}) for $\ln
\rho$, $s$ and $U_Z$. In addition we assume $U_Z=0$ at $R=R_0$.
Thus, we have in the star buffer zone
\begin{eqnarray}
&&\frac{{\rm D} \ln \rho(R)}{{\rm D} t}=...-\frac{\ln \rho(R)-\ln
\rho(2R_0-R)}{\tau}\lambda(R),
\\
&&\frac{{\rm D} U_{Z}(R)}{{\rm D}
t}=...-\frac{U_{Z}(R)-U_{Z}(2R_0-R)}{\tau}\lambda(R),
\\
&&\frac{{\rm D} s(R)}{{\rm D}
t}=...-\frac{s(R)-s(2R_0-R)}{\tau}\lambda(R).
\end{eqnarray}

Since the main goal of this paper is to consider the gas flow in the
vicinity of the equatorial point $E$, we consider a domain located
inside the disk with a vertical size smaller than the height
where the gas becomes cold and optically thin. To imitate a disk
photosphere in the surface buffer zone we include an
additional cooling term to create a vertical temperature
gradient and to allow gas to escape through the surface boundary.

First, we attempt to consider a non-rotating star.
It turns out that in this case the initial distribution of the main quantities
(temperature, density, and velocity) have to be close to the
final state, because otherwise inhomogeneities in the temperature result
in large radiation pressure gradients which, in turn, generate
large local velocities.
Such velocity perturbations may produce a local decrease of
density, which will lead to a decrease in radiative cooling, and
hence to an increase in temperature. The resulting radiation
pressure gradient will decrease the density even
further, which leads therefore to an instability.  (Note that this
does not happen in the case of a smaller gas temperature because
then the gas pressure dominates over the radiation pressure.)

Finding suitable initial conditions is a difficult task. It turns out that
it is easier to consider first a rotating star, and then to
decrease its rotational velocity down to the necessary value.
However, the final rotational velocity still has to be
considerable so that centrifugal and gravitational forces
are of the same order of magnitude. In the opposite case,
i.e.\ when the star is almost non-rotating, the gravity force has to be
balanced by the radiation pressure gradient, which, in turn,
should be negative near the stellar surface. However, it is not
clear how to realize this, because the main heating mechanism is due
to viscous friction which is maximum in the middle of the
boundary layer rather than at the stellar surface.

At the current stage we assume that the rotational velocity of
the star corresponds to the Keplerian velocity at the
stellar radius (i.e.\ $\alpha=1$). Also, we take a uniform initial
distribution of temperature in the $R$ direction. In that case, at
the beginning of the calculation, the huge gravitational
force near the stellar surface is balanced by the centrifugal force.
Depending on the temperature gradient, the $R$-component of the
radiation pressure force nearly vanishes and the
resulting $R$ component of the velocity appears to be small.

While the $Z$ component of the gravitational force is much smaller
than the $R$ component, its influence on the gas motion in the $Z$
direction is crucial and leads to strong flow of gas into the
domain through the surface boundary and to an accumulation of gas
near the equatorial point $E$. To avoid numerical problems, we
assume that initially the $Z$ component of gravitational force is
balanced by the corresponding component of the radiation pressure
force. Therefore, the initial distribution of temperature takes
the form
\begin{equation}
T(R_i)=T(R_{i-1})-\frac{3 G M_{\rm star}c \rho(R_i) Z_{i} }{16
\sigma_{\rm SB}R_{i}^3 T^3(R_i)} \Delta R, \;\; i=1..N_R,
\end{equation}
where $\Delta R$ is the size of the mesh in the $R$ direction.
We use the same initial density distribution as in the
gas pressure dominated case (see Sec.~\ref{sec:gpd}).

In Fig.~\ref{fig:result2} we present temperature and velocity
fields (see also the sketch in Fig.~\ref{fig:geom}).
Here we show temperature
rather than density (as was done in Fig.~\ref{fig:result1}), because in
the radiation pressure dominated case the flow of the gas is mostly
determined by the radiation pressure force and hence by the
temperature distribution.
In addition, we take a larger
domain size because the thickness of the boundary layer now
appears to be an order of magnitude larger than that in
Sec.~\ref{sec:gpd}.

The results of the calculation are also presented in
Fig.~\ref{fig:result2_2}, where density, temperature, and
velocity of the gas are shown as functions of $R$ for three
different distances from the mid-plane, $Z=10$, 50  and 110 m. We find that at
higher latitudes the gas rotates with a velocity that is comparable
to the rotational velocity of the star, while in the equatorial
plane its rotational velocity is smaller than the stellar surface speed.
This means that in the equatorial plane the centrifugal force is larger
than the gravitational force, while at the higher latitudes the
centrifugal force is smaller than the gravitational force.
Such a relation between the main forces
would result in accretion in the equatorial plane and excretion at
higher latitudes, provided the gas pressure was much larger than the
radiation pressure. However, since now the radiation pressure
dominates, we obtain the opposite result: the gas accretes only at
higher latitudes, while in the equatorial plane it is excreting.

In Fig.~\ref{fig:forces} we present, as a function of $R$,
the sum of the $R$ components
of gravitational, centrifugal, and radiation pressure forces,
$F_{\rm tot}=F_{\rm gr}+U_{\Phi}^2/R+\kappa\ \fradscal/ c$.
One can see that at a higher latitudes
the total force is at some radius negative, $F_{\rm tot} <0$,
so the generated radial velocity is negative ($U_{R}<0$),
which means accretion, while in the equatorial plane $F_{\rm
tot} >0$ and $U_{R} >0 $, so the gas is excreting. The dominant
role of the radiation pressure in driving the velocity field is
also clear from analyzing the temperature in
Fig.~\ref{fig:result2}. One can see that the temperature decreases
outward in the equatorial plane and increases outward at $Z
\simeq 110$~m near the disk buffer zone. Note that along the
stellar surface the temperature is almost constant, so the
$Z$ component of the gravitational force dominates here, and
causes the gas to sink toward the equatorial plane.

\section{Conclusions}

We have studied the gas flow in close vicinity of a neutron star in
a low mass X-ray binary and have assumed that the
magnetic field is negligible.
The main purpose of this work was to investigate the flow near the
equatorial plane between disk and star,
so the curvature of the stellar surface in the latitudinal direction
was neglected and cylindrical coordinates were used.

In the unrealistic, gas pressure dominated case
the gas temperature is  $T \simeq 5 \times 10^6$ K (which is
about an order of magnitude smaller than the observed value).
If the star does not rotate, the gravitational force at a radius close to the
stellar surface should be balanced by the gas pressure force. To have
gas pressure and gravitational forces of the same order of
magnitude, the stellar mass was chosen to be about two orders of magnitude smaller
than the real mass of a neutron star. In this case the maximum
release of energy occurs in the middle of the boundary layer,
where the gas velocity gradient (and hence the viscous heating)
reaches a maximum, while the radiation pressure force at the
stellar surface is directed inward.

For a realistic neutron star mass, $M_{\rm star} =1.4\msun$,
the gas pressure gradient at the
stellar surface becomes negligible compared with the gravitational force.
The latter is balanced by the radiation pressure,
so the gas temperature is about $T \simeq 10^8$ K.
Unlike the gas pressure dominated case, the radiation pressure force
is directed outward rather than inward.
Thus, the maximum energy release occurs directly at the stellar surface.
However, it is not clear
how to realize such a scenario, where the gas is heated by
viscous friction between the differentially rotating gas layers.

The picture becomes crucially different if one
considers a rotating neutron star: the gravitational force can now
be balanced by the centrifugal force.
Here we have assumed that the neutron star rotates with the Keplerian
velocity at the stellar radius. Alternatively, one might find a
solution for smaller rotational velocities by gradually decreasing
it down to the required value, using the result of
calculating it for a larger velocity as an initial approximation
for calculation with the smaller value.

We find that at higher latitudes the centrifugal force is
larger than the gravitational force, while at the equatorial plane the
gas rotates with a velocity that is considerably smaller than
the corresponding Keplerian value. It would be reasonable to assume
that the gas is accreting near the equator and excreting at
higher latitudes. However, we find the opposite: the accretion
occurs only at higher latitudes, while in the equatorial plane
the gas is excreting. This is related to the temperature
distribution, and therefore, to the radiation pressure force,
which is now dominant. We find that near the equatorial plane the
temperature decreases outward, so the gas is pushed
away from the surface by radiation pressure. At higher latitudes,
some distance away from the surface, the temperature decreases
inward, resulting in accretion. The circulation of the gas
is closed by a flow along the stellar surface from high
to low latitudes, because the temperature is almost
constant in this direction and the gas flow is  controlled only by
the tangential component of gravity.

Finally, one should note that, since we have considered only a
laminar two-dimensional model of the
boundary layer at the neutron star surface, we have assumed that the
turbulent viscosity is constant everywhere and that it can be treated as an
input parameter. Future three-dimensional simulations will allow us to model
turbulent processes more accurately.
However, even the results of the two-dimensional
simulations give us some clues for understanding the physical
processes near the neutron star surface. These results can
in principle be used for a more detailed description of the vertical
structure of the boundary layer and for calculating spectra
of neutron star radiation. Furthermore, the presented results may be
useful for understanding the nature of quasi-periodic oscillations.

\section*{Acknowledgments}
This work was supported by the Academy of Finland grant 110792 and
the Magnus Ehrnrooth Foundation.
We acknowledge the allocation of computing resources provided by the
Center for Scientific Computing in Finland.

\newcommand{\yana}[3]{ #1, {A\&A,} {#2}, #3}
\newcommand{\ymn}[3]{ #1, {MNRAS,} {#2}, #3}
\newcommand{\ypasp}[3]{ #1, {PASP,} {#2}, #3}
\newcommand{\yapj}[3]{ #1, {ApJ,} {#2}, #3}
\newcommand{\yjour}[4]{ #1, {#2}, {#3}, #4}
\newcommand{\yproc}[5]{ #1, in {#3}, ed.\ #4 (#5), #2}

\label{lastpage}
\end{document}